\documentclass[twocolumn, pra, floatfix, superscriptaddress, aps, showpacs]{revtex4-1}
\usepackage{graphicx}

\begin{document}

\title{Calorimetry of a harmonically trapped Bose gas}

\author{S.~K.~Ruddell}\email{sam.ruddell@auckland.ac.nz}\affiliation{The Dodd-Walls Centre for Photonic and Quantum Technologies}\affiliation{Department of Physics, University of Auckland, Private Bag 92019, Auckland, New Zealand}

\author{D.~H.~White}\affiliation{The Dodd-Walls Centre for Photonic and Quantum Technologies}\affiliation{Department of Physics, University of Auckland, Private Bag 92019, Auckland, New Zealand}

\author{A.~Ullah}
\altaffiliation[Present address: ]{Department of Physics, University of Malakand, Khyber Pakhtunkhwa, Pakistan}\affiliation{Department of Physics, University of Auckland, Private Bag 92019, Auckland, New Zealand}

\author{D.~Baillie}\affiliation{The Dodd-Walls Centre for Photonic and Quantum Technologies}\affiliation{Department of Physics, University of Otago, Dunedin, New Zealand}

\author{M.~D.~Hoogerland}\affiliation{The Dodd-Walls Centre for Photonic and Quantum Technologies}\affiliation{Department of Physics, University of Auckland, Private Bag 92019, Auckland, New Zealand}

\begin{abstract}We experimentally study the energy-temperature relationship of a harmonically trapped Bose-Einstein condensate by transferring a known quantity of energy to the condensate and measuring the resulting temperature change. We consider two methods of heat transfer, the first using a free expansion under gravity and the second using an optical standing wave to diffract the atoms in the potential. We investigate the effect of interactions on the thermodynamics and compare our results to various finite temperature theories.
\end{abstract}

\pacs{64.70.Tg, 05.30.Rt, 67.85.Hj}
\maketitle

\section{Introduction}

Calorimetric studies have long been valuable tools for rigorous tests of physical law, ranging from Joseph Black's early work on latent heat, to measurements of the metallic specific heat showing the inadequacy of Drude's classical model. More recently, the high degree of control available in cold atomic gases has opened up exciting avenues for experimental verification of finite temperature theories of Bose and Fermi gases \citep{Blakie2007}. To date, however, there have been few works experimentally investigating the energy-temperature relationship of a harmonically trapped Bose gas. Pioneering work performed by Ensher~\emph{et al.} relied on extracting both release energy and temperature information from time-of-flight images at different evaporation points \cite{Ensher1996}. This work was extended by Gerbier~\emph{et al.}, whose measurements of the release energy were found to be in good agreement with Hartree-Fock theory for an interacting gas \cite{Gerbier2004}. Gati~\emph{et al.} have measured temperature dependent phase fluctuations of an ideal Bose gas and revealed qualitatively that the system deviates from a classical gas \cite{Gati2006}.

To further study the energy dependence of Bose-Einstein condensate (BEC) thermodynamics, two options present themselves. Firstly, an improvement in the ability to extract thermodynamic information from time-of-flight images could extend the results presented in \cite{Ensher1996,Gerbier2004}. Secondly, an alternative method to study the energy-temperature relationship could be envisioned, allowing the total internal energy of the system to be measured, rather than just the release energy. New methods to study this relationship are emerging, motivated by the recent characterization of the heat capacity of a strongly interacting Fermi gas \cite{Kinast2005,Ku2012}. An attempt to measure the specific heat of an ultracold Bose gas using a time-dependent trapping potential, as well as by heating using laser pulses, has been performed \cite{deJong2013}, although obtaining accurate data was found to be impractical in their system. A recent experiment has extracted information regarding the heat capacity of a Bose gas using global variables \cite{Shiozaki2014}.

In this paper we follow the theoretical proposal of Blakie \emph{et al.} to transfer a known quantity of irreversible work to a BEC and measure the resulting temperature~\cite{Blakie2007}. By utilizing two independent methods we perform known, precise amounts of work on a$~^{87}$Rb condensate. The resulting temperature is measured after a period of thermalization, giving the transferred energy as a function of temperature. This provides a rigorous test of the energy dependence of the thermodynamics of our system, as energy and temperature measurements are performed independently. Our approach contrasts that of Ensher \emph{et al.}, who measure their system at differing evaporation points, extracting both energy and temperature information from time-of-flight images. Furthermore, our approach is not sensitive to the ground state energy of the system present at $T=0$, allowing a more direct comparison with the specific heat. Our results from the two methods compare well, both with each other and with Hartree-Fock numerical calculations for an interacting gas.

The paper is organised as follows: Section \ref{sec:exp} reviews the experimental parameters and methods. Sec.~\ref{sec:temp} details the temperature measurements.
The two methods of energy transfer are discussed in Sec.~\ref{sec:grav} and Sec.~\ref{sec:kick}. We close with a discussion and conclusion in Sec.~\ref{sec:dis} and Sec.~\ref{sec:con} respectively.

\section{Experimental parameters}~\label{sec:exp} 
\begin{figure}[t]
\centering
\includegraphics[width=83mm]{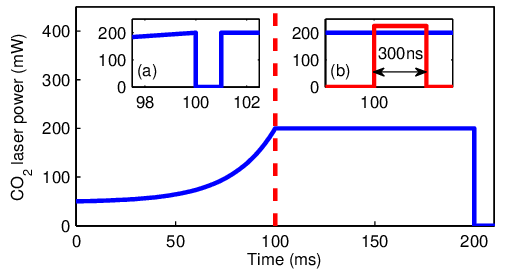}
\caption{(Color online) The trap laser power sequence in the experiment. Following the production of the BEC, the laser power is adiabatically ramped up with an exponential profile over 100 ms, increasing the trap depth to $3.3~\mu\mathrm{K}$ - $5~\mu\mathrm{K}$. Work is then done on the condensate in one of two ways: (a) the release of the atoms for a time $t_{heat} =$ 0-1000~$\mu$s leads to falling expansion of the cloud, resulting in increased kinetic and potential energy when the trap is subsequently reinstated; (b) a 300 ns pulse of an off-resonant standing wave leads to diffraction of a fraction of the atoms. Following either of these is a 100 ms period of thermalization, before the condensate is left to expand for 10 ms to allow the momentum distribution to be imaged via time-of-flight.}
\label{fig:1}
\end{figure}

Our experiment involves a BEC of \mbox{$\sim2\times 10^4~^{87}$Rb} atoms prepared in the \mbox{$\left|F=1;m_F=-1\right\rangle$} ground state, and held in an optical dipole trap~\cite{Wenas2008}. The trap is formed at the intersection of two focused CO$_2$ laser beams, with wavelength 10.6~$\mu$m, and each with a $1/e^2$ radius of 33~$\mu$m. The CO$_2$ laser power is stabilised using a closed-loop feedback system to ensure long term reproducibility of the trap depth, BEC atom number, and temperature. After loading atoms into the dipole trap from a magneto-optical trap operating on the 780.2~nm $(5s)^2S_{1/2} \rightarrow (5p)^2P_{3/2}$ transition, a 6~second evaporative cooling sequence is used to produce a BEC. We then execute the experimental sequence shown in Figure \ref{fig:1}. The laser power is adiabatically ramped to a higher value over 100~ms using an exponential profile. The deeper potential resulting from this ramp prevents atom loss during the heating process. The adiabaticity of this ramp has been confirmed by ensuring that negligible non-condensed fraction exists following the ramp and a 100~ms hold time at the final laser power.
We then transfer a precise amount of energy to the system, using one of two methods, before allowing the system to rethermalize for 100~ms. 

We approximate our optical dipole trap as a harmonic potential characterized by a set of frequencies $\omega_j$ that define the potential in three dimensions. These frequencies are measured through a parametric heating process \cite{Friebel1998}, where the trap depth is modulated sinusoidally for a period of 200~ms with an amplitude of $\sim10$\%
of the total trap depth. Parametric excitation along dimension $j$ occurs for \mbox{$\omega_{mod} = 2\omega_j/n$}, for integer $n$. Measurements of the excitation frequencies allow us to characterize our trap and calculate the critical condensation temperature for an ideal Bose gas $T_c^0$, given by

\begin{equation}
	T_c^0 = \frac{\hbar\bar\omega}{k_B}\left[\frac{N}{\zeta(3)}\right]^{1/3},
\end{equation}
where $\bar\omega=(\omega_x\omega_y\omega_z)^{1/3}$ is the geometric mean of the trapping frequencies, $k_B$ is the Boltzmann constant, $N$ is the number of atoms, and \mbox{$\zeta(\alpha) = \sum_{n=1}^{\infty}n^{-\alpha}$} is the Riemann zeta function (e.g. see \cite{Giorgini1996}).

We ensure that the initial system is at zero temperature by evaporating to a point where the thermal fraction can no longer be observed. We find that any further lowering of the trapping potential only leads to strong depletion of the condensate, and conclude that the zero initial temperature condition is satisfied.

In our system, the interaction energy of the initial BEC far outweighs the kinetic energy, as \mbox{$Na_s/a_{ho}\gg1$}, where $a_s$ is the s-wave scattering length, and \mbox{$a_{ho} = \sqrt{\hbar/m\bar\omega}$} is the characteristic harmonic oscillator length~\cite{Giorgini1997}, where $m$ is the mass of an atom. Typically we find that \mbox{$Na_s/a_{ho}> 100$}, and can assume the Thomas-Fermi approximation applies.

\section{Temperature measurement}
\label{sec:temp} The temperature and number of atoms are measured using time-of-flight imaging with resonant absorption, and the properties of the atomic clouds are inferred from these images. Following an experimental sequence, the dipole trap containing the atoms is rapidly switched off using an acousto-optic modulator, and the atoms are allowed to freely expand for 10~ms. After a repumping pulse, the atoms are probed with a $100~\mu \mathrm{s}$ pulse on resonance with the \mbox{$\left|F=2\right\rangle \rightarrow\left|F^\prime=3\right\rangle$} transition. The probe light has an intensity of \mbox{$1~\mathrm{mW/cm^2}$}, which is less than the saturation intensity of \mbox{$1.6~\mathrm{mW/cm^2}$}.

When processing time-of-flight images we first subtract the average background of all images and apply a fringe-removal algorithm \cite{Ockeloen2010}, improving our signal-to-noise ratio and ability to detect low density components of the expanded atomic cloud. The images are then integrated along the $x$- and $y$-dimensions to obtain two one-dimensional density profiles, from which we extract both the temperature and the number of atoms of our sample.

Above the critical temperature, the expansion of an ideal Bose gas evolves according to a simple scaling relation, where we can define the effective temperature after an expansion time $t$ in dimension $i = x,y$ as 
\begin{equation}
T_i = \frac{m}{k_B}\frac{\omega_i^2\sigma_i^2(t)}{1+(\omega_it)^2}.
\end{equation} 
Here, $\sigma_i^2(t)$ is the variance of the resulting distribution as a function of the expansion time. Far above the experimentally observed critical temperature $T_c$, this can be determined by a fit to a Gaussian function. Close to $T_c$ and below, the density distribution of the thermal cloud becomes predominantly the Bose-distribution, and by setting the chemical potential to zero, it can be described by a Bose-enhanced Gaussian \cite{Ketterle1999}. In the hydrodynamic regime, there is the possibility of anisotropic expansion for a very elongated trap, which occurs when the mean free path of the atoms is less than the dimension of the trap (e.g. see \cite{Gerbier2004}). For the experiments presented in this paper our trap is nearly isotropic, as we have $\omega_x \approx 1.4\omega_y$, with $\bar\omega/(2\pi) = (220\pm5)~\mathrm{Hz}$ and $(271\pm5)~\mathrm{Hz}$ for the two experiments. In addition, the 10~ms expansion time used sets $(\omega_it)^2\gg1$, and hence we can assume that the expansion of the ideal thermal component to be isotropic, as observed experimentally. We therefore are able to assume that $T_x = T_y = T$.
 
\begin{figure}[t]
\centering
\includegraphics[width=83mm]{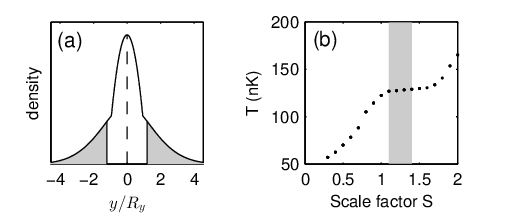}
\caption{An example temperature measurement below $T_c$. (a) The bimodal fit applied to a sample along the $y$-dimension, where we have inferred the temperature from a fit to the grey shaded region, here with $S=1.2$. Here the position $y$ has been scaled by the Thomas-Fermi radius $R_y$. (b) The effect of scanning $S$ on the temperature measurement for a sample image. The grey shaded region indicates the typical values of $S$ used to infer the temperature.}
\label{fig:splot}
\end{figure}

Below the critical temperature there exists a non-negligible condensed fraction, and as such the assumption of a ballistic expansion for the entire cloud is no longer valid. We therefore extract the temperature using the method presented in \cite{Szczepkowski2009}. 
The central interacting region is systematically excluded from our measurements by performing multiple fits of a Bose-enhanced Gaussian to the wings of the profile, with a varying cut-off width for the excluded central region. For each fit, the width of this region is determined by a scaling factor $S$, such that the region $|x_i|\leq SR_i$ is excluded from the fit, as shown in Figure \ref{fig:splot}(a). Here $R_i$ is the Thomas-Fermi radius of the condensed fraction in dimension $x_i$, with $i=x,y$. $S$ is chosen such that the region excluded for the fit is larger than the width of the condensed fraction, as sampling this region would cause us to systematically underestimate the temperature. On the other hand, if $S$ is too large, we are limited by the signal-to-noise ratio of our images. There exists an intermediate region where the measured temperature depends only weakly on the width of the excluded region, which we typically find to be $1.1\leq S \leq 1.4$. We infer the temperature from this region, as illustrated in Figure \ref{fig:splot}(b).

This technique is only valid when the extent of the thermal profile is much larger than that of the condensed fraction, and thus fails at very low temperatures. In our experiment this is more pronounced in one direction, due to higher oscillator frequency in that dimension ($\omega_x \approx 1.4\omega_y$). We therefore choose to extract temperature information from the $y-$dimension, where the resulting momentum distribution of the condensate fraction is narrower, and apply a temperature cut-off at $T\leq 0.3T_c$. An investigation of the very low temperature region ($T<0.3T_c$) would require a new thermometry technique to be used, such as recently presented by Olf~\emph{et al.} This technique involves analyzing the decoherence of a quantum superposition of spin states, which allows them to measure temperatures as low as $0.02T_c$ \cite{Olf2015}. 

To check the accuracy of our temperature measurement well below $T_c$, we have simulated the mean-field effect of the condensate on the thermal cloud as the gas expands. We model the expansion of the condensate using hydrodynamic scaling \cite{Castin1996}. For the thermal cloud we use two approaches, Monte Carlo with $10^7$ test particles \cite{Jackson2002}, and a scaling approach after \cite{Hu2003}. In both cases we find that the effect on the temperature measurement is less than 10\%.

For higher temperatures, still below $T_c$, the magnitude of this effect is decreased due to the smaller condensed fraction. We estimate the uncertainty on $T$ here to be 5\%, accounting for the uncertainty in length calibration, possible collisional effects during expansion, and any variance of the temperature measurement depending on the specific choice of $S$.

The atom number is obtained from a bimodal fit to the absorption profile \cite{Ketterle1999}. For temperatures below $T_c$, we perform a bimodal fit of a Bose-enhanced Gaussian to the thermal fraction, and a Thomas-Fermi profile to the condensed fraction using the method presented in \cite{Szczepkowski2009}. Integration over the entire bimodal profile gives the total optical density $\sum n_{OD}$, which is directly proportional to the number of atoms $N$, given by $N =\sum n_{OD}A_p/\sigma$. Here $A_p$ is the area of a pixel, and $\sigma = \alpha_s\sigma_0$ is the experimental absorption cross-section. Here $\sigma_0=3\lambda_P/(2\pi)$, with $\lambda_P$ being the wavelength of the probe laser beam used in the imaging process, and $\alpha_s$ is a scaling factor that allows us to account for the Clebsch-Gordan coefficients combined with the experimental distribution of magnetic substates following repumping. For an even distribution across the magnetic substates, as would be the case after repumping, $\alpha_s=0.47$. Experimentally, we independently determine $\alpha_s$ by observing experimental images having a temperature close to $T_c$. We then scale our measured number of atoms such that the measured temperature agrees with the theoretical critical temperature $T_{int}$, which we have determined to be \mbox{$T_{int} = 0.94T_c^0$} from Hartree-Fock numerical simulations. Due to the nature of the interacting transition, the condensed fraction does not go abruptly to zero as the temperature crosses $T_{int}$, as in the ideal case. Using this method we determine that $\alpha_s = 0.45\pm 0.07$.

\section{Energy transfer via gravity and expansion.}\label{sec:grav}~We utilize two separate methods for transferring energy to our system. The first method was proposed by \mbox{Blakie {\it et al.} \cite{Blakie2007}}. Here we consider an irreversible work process: a Bose-Einstein condensate is released from the trapping potential and allowed to expand under the influence of gravity. After a time $t_{heat}$ (typically $0-1000~\mathrm{\mu s}$), the atoms are recaptured, and allowed to rethermalize. 

There are three contributions to the amount of work done on the atoms during this process. One, the atoms fall under gravity (acceleration $g$) and gain kinetic energy; two, the displacement $h=\frac{1}{2}gt_{heat}^2$ from the fall leads to a potential energy gain when the trap is reinstated; three, the larger cloud size after the expansion results in greater potential energy when the trap potential is restored. Energy from the first two contributions will be coupled to a center-of-mass oscillation in the potential in the $z$-direction, known as the ``Kohn'' mode \cite{Dobson1994}. Although this mode will theoretically persist in a harmonic trap, we observe that these oscillations are damped in the thermalization process after $\sim50$~ms. We attribute this to anharmonicities in the Gaussian laser trap profile \cite{Pantel2012}. Due to the observed damping of the Kohn mode, we consider this energy to be completely available for rethermalization, and to have the form

\begin{equation}
\label{eq:Edrop}
\mathcal{E}_{drop} = N\left(\frac{1}{2}m\omega_z^2h^2 + mgh\right),
\end{equation}
\noindent where $\omega_z$ is the trap frequency parallel to the direction of gravity.

\begin{figure}[t]
\centering
\includegraphics[width = 83mm]{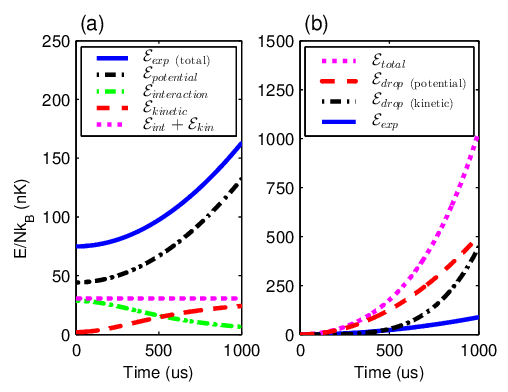}
\caption{(Color online) Simulation results showing the energy per atom for a mean-field expansion of the ground state of our harmonic trap as a function of $t_{heat}$. (a) The energy contributions due to a free expansion without gravity about the trap minimum, including the ground-state energy. $\mathcal{E}_{exp}$ is the sum of the potential, interaction, and kinetic terms. (b) The contributions to the total energy transferred to the system, excluding any initial ground state energy. Here $\mathcal{E}_{exp}$ is the same curve shown in (a), minus the initial ground state energy, and $\mathcal{E}_{drop}$ has been split into its potential and kinetic components.}
\label{fig:simTest}

\end{figure}

\begin{figure}[t]
\centering
\includegraphics[width = 83mm]{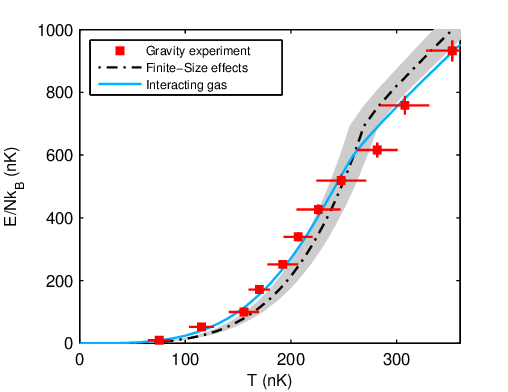}
\caption{(Color online) Experimental data for the gravity experiment plotted with theoretical curves. Experimental parameters are $N = (2.2\pm0.3)\times 10^4$ and $\bar\omega/2\pi = 220\pm5$~Hz. The interacting gas curve is a Hartree-Fock numerical simulation, with the $T=0$ ground state energy subtracted, to represent the transferred energy. The shaded region represents our uncertainty in determining $T_c^0$ for the finite-size effects theoretical curve.}
\label{fig:2}
\end{figure}

To calculate the energy acquired from expansion of the cloud, we assume that the atoms will undergo a self-similar expansion, with widths evolving according to \mbox{$R_j(t) = \lambda_j(t)R_j(0)$} \cite{Castin1996}. Here $R_j(t)$ are the Thomas-Fermi profile condensate widths in directions $j = x,y,z$, and the evolution of $\lambda_j$ is given by

\begin{equation}
\ddot\lambda_j = \frac{\omega^2_j(0)}{\lambda_j\lambda_x\lambda_y\lambda_z},
\end{equation}
\noindent with $\lambda_j(0) = 1$. Blakie {\it et al.}~\cite{Blakie2007} have shown that the energy transferred to the system due to a symmetric expansion about the trap minimum is given by
\begin{equation}\label{eq:Blakie1}
\mathcal{E}_{exp}=\frac{N\mu_{TF}}{7}\left(2-5\bar\gamma^{6/5}+\sum_{j=1}^3\gamma_j^2 \lambda_j^2(t_{heat})\right),
\end{equation}
\noindent where $\mu_{TF}$ is the Thomas-Fermi chemical potential, \mbox{$\gamma_j = \omega^\prime_j/\omega_j$} is the ratio of trapping frequencies before and after $t_{heat}$, and \mbox{$\bar{\gamma} = \left(\gamma_x\gamma_y\gamma_z\right)^{1/3}$}. In our experiment we constrain the trapping frequencies both before and after $t_{heat}$ to be identical, such that $\gamma_j=1$, reducing this expression to
\begin{equation}\label{eq:Blakie2}
\mathcal{E}_{exp}=\frac{N\mu_{TF}}{7}\left(-3+\sum_{j=1}^3 \lambda_j^2(t_{heat})\right).
\end{equation} The various contributions to $\mathcal{E}_{exp}$ are shown in Figure \ref{fig:simTest}(a).
The total amount of energy transferred to the system and available for rethermalization is then the sum of the contributions $\mathcal{E}_{drop}$ and $\mathcal{E}_{exp}$. 

We wish to reduce the dependence of our energy calculation on the absolute number of atoms $N$, which has an uncertainty of 15\% due to the error in $\alpha_s$. We therefore choose to calculate the energy per particle, rather than the total energy transferred to the atoms. In this scenario, only $\mathcal{E}_{exp}/N$ maintains a dependence on $N$, with $\mu_{TF}$ proportional to $N^{2/5}$. This term accounts for less than 20\% of the total energy in our experiment, as shown by a numerical simulation in Figure \ref{fig:simTest}. Here we have calculated the ground state of our harmonic potential in three dimensions by solving for the ground state of the Gross-Pitaevskii equation for our trap parameters. This initial condition is then allowed to expand under gravity for up to 1000~$\mu$s using a split-step Fourier method.

\begin{figure}[t]
\centering
\includegraphics[width = 83mm]{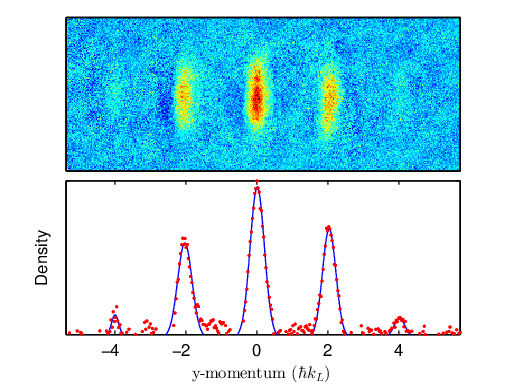}
\caption{(Color online) Experimental determination of the energy of a kick. The top image represents a time-of-flight image. The bottom image shows the integral of the time-of-flight image along the $x$-dimension (vertical in image), allowing us to determine the fraction of atoms at each momenta by fitting a Gaussian profile to each peak.}
\label{fig:kickTest}
\end{figure}

The uncertainty in $\mathcal{E}_{drop}/N$ is mainly due to the uncertainty in the measurement of the trap frequency $\omega_z$, used for determining the potential energy contribution in equation \ref{eq:Edrop}, giving an uncertainty of 4\% for this component. The accuracy of the kinetic component of $\mathcal{E}_{drop}/N$ is due to our accuracy in determining the local value of $g$, as well as the timing accuracy in our experiment. This component accounts for more than half of the total energy transferred to the system, and has an uncertainty \mbox{of $<1$\%}. 

Experimental results are shown in Figure \ref{fig:2}, and have been plotted with a Hartree-Fock calculation for this system. As we are only interested in the energy transferred to the system, the ground state energy present at $T=0$ is subtracted from the theoretical curve. This allows us to make close comparisons with the specific heat of the system, which is defined as the temperature derivative of the energy per particle. The theoretical curve for an ideal gas with finite-size effects includes a shaded region representing the uncertainty in our determination of the critical temperature $T_c^0$, which is mainly due to the uncertainty in the absolute measurement of $N$. Notwithstanding the 15\% error in $N$, the Hartree-Fock numerical simulation gives a better description of the behavior of our data than an ideal gas having finite-size effects.

\section{Energy transfer via an optical phase grating}\label{sec:kick}~To support our previous evidence, which relies on a calculation of the work done on the BEC, we utilize a separate method which allows for a direct measurement of the transferred energy. Here, we transfer energy to the system using a single pulse of an optical standing wave, as indicated in Figure \ref{fig:1}(b), before allowing the system to rethermalize. Using a setup analogous to the atom-optics kicked rotor pioneered by Raizen and co-workers \cite{Moore1995,Raizen1999}, we apply a short 300 ns pulse to the atoms, diffracting the system into quantized momentum orders. An additional advantage of this method is that the Kohn mode is naturally not present, as the diffraction is symmetric about zero momentum. We use a pair of counter-propagating laser beams to form our standing wave, with the beams red-detuned by $120$~GHz from the \mbox{$\left|F=1\right\rangle \rightarrow\left|F^\prime=2\right\rangle$} resonant transition, such that the effects of spontaneous emission are negligible. The atoms are diffracted into quantized momentum states, with the $n$th momentum state having momenta $2n\hbar k_L$, where $k_L$ is the wavenumber of the laser and $n$ is integer.

\begin{figure}[t]
\centering
\includegraphics[width = 83mm]{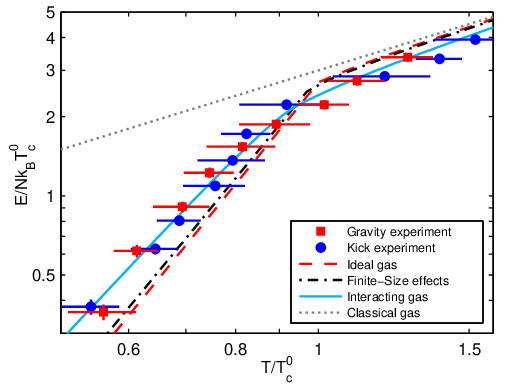}
\caption{(Color online) Logarithmic plot of experimental data and theoretical curves. The theoretical curves represent energy transferred to the system only, and any ground state energy at  $T=0$ has been neglected. Parameters for the gravity experiment are $N = (2.2\pm0.3)\times 10^4$, $\bar\omega/2\pi = 220\pm5~\mathrm{Hz}$ and $t_{heat} = 0-1000~\mu$s. Parameters for the kicking experiment are $N = (2.1\pm0.3)\times 10^4$, $\bar\omega/2\pi = 271\pm5~\mathrm{Hz}$ and $k = 0\rightarrow1.9$.}
\label{fig:3}
\end{figure}

In a separate calibration experiment, we quantitatively measure the amount of energy transferred as a function of kick-strength, given by $k = \tau\Omega^2/\delta$, where $\tau$ is the pulse length, $\Omega$ is the Rabi frequency of a single beam and $\delta$ is the detuning. Experimentally, $k$ is varied by controlling the intensity of the laser for each pulse using an acousto-optic modulator, and we utilise $k = 0\rightarrow1.9$. We measure the resulting momentum distribution by turning off the trap immediately after the kick and observing the atoms after a 10~ms expansion time, as shown in Figure \ref{fig:kickTest}. The total energy transferred to the system is computed as
\begin{equation}
\mathcal{E}_{kick} = \frac{2N\hbar^2 k_L^2}{m}\sum_n f_n n^2,
\end{equation}
\noindent where $f_n$ is the fraction of atoms in the $n$th momentum state, for integer $n$. We again scale this energy measurement by $N$ to remove the dependence of our energy measurement on our determination of the absolute number of atoms. The uncertainty in $\mathcal{E}_{kick}/N$ is then due to shot-to-shot variation in intensity of the kicking laser and the temporal width of the pulse. We perform multiple calibration runs to obtain an average transferred energy, and allow the variation to be experimentally represented as a variation in the resulting temperature after thermalization. We find that the shot-to-shot variation in $\mathcal{E}_{kick}/N$ can be up to 10\%.

Once $\mathcal{E}_{kick}/N$ has been calibrated we repeat the experiment, but allow the atoms to rethermalize in the trap for 100~ms following the kick, before imaging the system via time-of-flight. The experimental data are shown in Figure \ref{fig:3}.

\begin{figure}[t]
\centering
\includegraphics[width = 83mm]{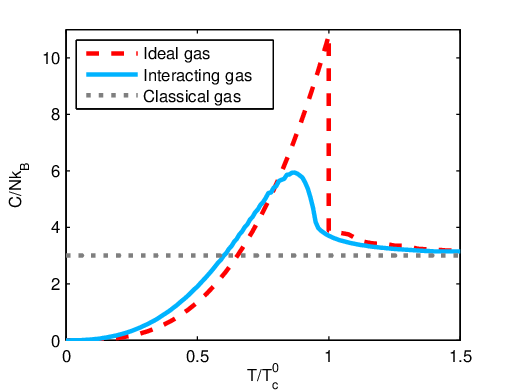}
\caption{(Color online) Theoretical curves showing the specific heat, defined as the temperature derivative of the energy per particle, for various theoretical treatments of a harmonically trapped Bose gas.}
\label{fig:hc}
\end{figure}

\section{Discussion}
\label{sec:dis}
Figure \ref{fig:3} shows both sets of experimental data on a logarithmic axis. Here we scale the temperature by the ideal gas critical temperature $T_c^0$, and the energy per particle by the characteristic energy of the transition $k_BT_c^0$. The data clearly deviates from the classical prediction of $E = 3Nk_BT$, where we would expect $E = k_BT/2$ in each of the three potential and three kinetic degrees of freedom from the equipartition theorem. There is also a deviation from the prediction for an ideal Bose gas. This deviation is beyond finite-size corrections \cite{Grossmann1995,Ketterle1996}, and is consistent with Hartree-Fock simulations of an interacting gas. Corrections to the Hartree-Fock approximation, such as in Hartree-Fock-Bogoliubov-Popov theory \cite{Giorgini1997}, are very small for our system, and impractical to measure. Theoretical studies have shown that the Hartree-Fock approximation can accurately reproduce the thermodynamic properties of a trapped Bose gas \cite{Krauth1996,Holzmann1999}, and we have confirmed through simulation that these two theories give very similar results for our system.

The specific heat is defined as the temperature derivative of the energy per particle, in our case with the external potential held constant. Taking numerical derivatives of our experimental data is impractical. We can instead make comparisons with the specific heat extracted from derivatives of the theoretical curves, shown in Figure \ref{fig:hc}. We find that our experiments support the notion that the presence of interactions will tend to increase the specific heat at low temperatures when compared to an ideal Bose gas. This can can be understood as a consequence of the repulsion of the thermal atoms by a large condensate fraction. The effective potential seen by the thermal atoms is modified to a ``Mexican hat'' type potential \cite{Tammuz2011}, increasing the volume occupied by the thermal atoms, thereby increasing the density of states. Consequently, this allows the otherwise ``saturated'' thermal cloud to hold more atoms, and hence more energy.


\section{Conclusion}
\label{sec:con}
We have directly measured the energy-temperature relationship of an interacting, harmonically trapped, ultracold Bose gas. Two separate calorimetric techniques have produced similar results; namely, that interactions lead to an increased specific heat from the ideal gas prediction, which is proportional to $T^3$ below $T_c^0$. We have performed quantitative measurements, utilising independent determinations of the energy and temperature, that are well described by Hartree-Fock theory. Future research could involve a thorough investigation of the effect of interactions on the specific heat by employing Feshbach resonances, an investigation into ways to reduce error in the experiment, and a detailed investigation of the thermalization process.

We are grateful to P.~B.~Blakie for useful discussions. This research was supported by the Marsden Fund, administered by the Royal Society of New Zealand on behalf of the New Zealand Government.
\\
\end{document}